\documentclass[11pt,twoside]{article}
\usepackage{./asp2014}

\aspSuppressVolSlug
\resetcounters

\begin{document}

\title{Statistics of magnetic field measurements in OB stars}
\author{A.F.~Kholtygin$^1$, S.~Fabrika$^2$, S.~Hubrig$^3$, G.A.Chuntonov$^2$, A.S. Medvedev$^2$, Yu.V.Milanova$^3$,
        M.~Schoeller$^4$, G.G.Valyavin$^2$, O.A.Tsiopa$^5$, S.V. Boronina$^1$ % and BOB collaboration  
\affil{$^1$Astronomy Department, Saint-Petersburg State University, Russia;    \email{afkholtygin@gmail.com}}
\affil{$^2$Special Astrophysical observatory, Nizhnii Arhyz, Russia;            } 
\affil{$^3$Leibniz-Institut f\"ur Astrophysik Potsdam (AIP), Potsdam, Germany;  }
\affil{$^4$European Southern Observatory, Garching, Germany;                    }  
\affil{$^5$Main Astronomical observatory, Saint-Petersburg, Russia;             }  
       }

% This section is for ADS Processing.  There must be one line per author.
\paperauthor{A.F.~Kholtygin}{afkholtygin@gmail.com}{}{Saint-Petersburg State University}{Aspronomyt}{Saint-Petersburg}{}{198504}{Russia}
\paperauthor{S.~Hubrig}{shubrig@aip.de}{}{Leibniz-Institut fur Astrophysik}{}{Potsdam}{}{14482}{Germany}
\paperauthor{M.~Schoeller}{mschoell@eso.org}{}{European Southern Observatory}{}{Garching}{}{85748}{Germany}
% \paperauthor{Sample~Author3}{Author3Email@email.edu}{ORCID_Or_Blank}{Author3 Institution}{Author3 Department}{City}{State/Province}{Postal Code}{Country}

\begin{abstract}
 We review the measurements of magnetic fields of OB stars and compile a catalog of magnetic OB stars. Based on 
available data we confirm that magnetic field values are distributed according to a log--normal law with a 
mean log(B)=2.53 and a standard deviation 
$\sigma=0.54$. We also investigate the formation of the magnetic field of OBA stars before the Main Sequence (MS). 

\end{abstract}

%==================================================================================================================================1
\section{Introduction}

The origin of magnetic fields in massive stars is still poorly known. Many authors argued that magnetic fields 
could be fossil, or 
it may be generated by a strong binary interaction in stellar mergers, or during a mass transfer or common 
envelope evolution.
The answers to questions related to the origin and evolution of magnetic fields in massive stars require therefore 
additional efforts in this field.

%----------------------------------------------------------------------------==2.
\section{Statistical Characteristics of Magnetic Fields} 

As the most convenient characteristic of the stellar magnetic field we use the {\it rms} magnetic field 
\begin{equation} 
\label{Eq.Baver}
\langle B\rangle = \sqrt{\frac{1}{n}\sum_{k=1}^{n} \left( B_z^k\right)^2} \, 
\end{equation} 
where we sum over all measured values of mean longitudinal magnetic fields $B_z^k$ for a given star. 
Here $k$ is the running number of the individual observation, $n$ is the total number of observations. \cite{Kholtygin-2010a} showed 
that  in the case of dipole field configuration 
the {\it rms} field value $\langle B\rangle$ depends weakly on the rotational phase $\phi$ of observations, 
the rotation axis inclination angle $i$, and the angle $\beta$ between the rotational axis and the axis of 
magnetic dipole. 
This conclusion holds for quadrupole and other field configurations.   
As a measure of reality of the measured field values we use the criterion~(4) presented by~\cite{Kholtygin-2010a}.  
This criterion is equivalent to the condition that the absolute value of the measured magnetic field $|B_z|$ is 
3 or more times larger than the error of the measurements at least for one field measurement.

%----------------------------------------------------------------------------==3.
\subsection{Distribution of Magnetic Fields}  

The analysis of a differential magnetic field distribution $f(\langle B\rangle)$ (the magnetic field function) introduced 
by~\cite{Fabrika-1997} is important for understanding the origin of stellar magnetic fields. 
The function $f(\langle B\rangle)$ is defined as follows: 
\begin{equation}
\label{Eq.MFFdeterm}
f({\langle B\rangle})  \approx \frac{N({\langle B\rangle},{\langle B\rangle} +\Delta {\langle B\rangle})}{N\Delta {\langle B\rangle}} \, , 
\end{equation}
where $N({\langle B\rangle},{\langle B\rangle}+\Delta {\langle B\rangle})$ is the number of stars in the interval 
$({\langle B\rangle},{\langle B\rangle}+\Delta {\langle B\rangle})$, $N$ is the total number of stars with real 
measured {\it rms} field $\langle B\rangle$. 
 At present, only for about one dozen of O-type stars the presence of a magnetic field is confirmed
using high- and low-resolution spectropolarimetry. For remaining 10 stars the presence of a weak
magnetic field is still under debate \citep{Hubrig-2011,Hubrig-2013}. 

%#########################################################################################################################==Figure1
\articlefigure[width=1.00\textwidth]{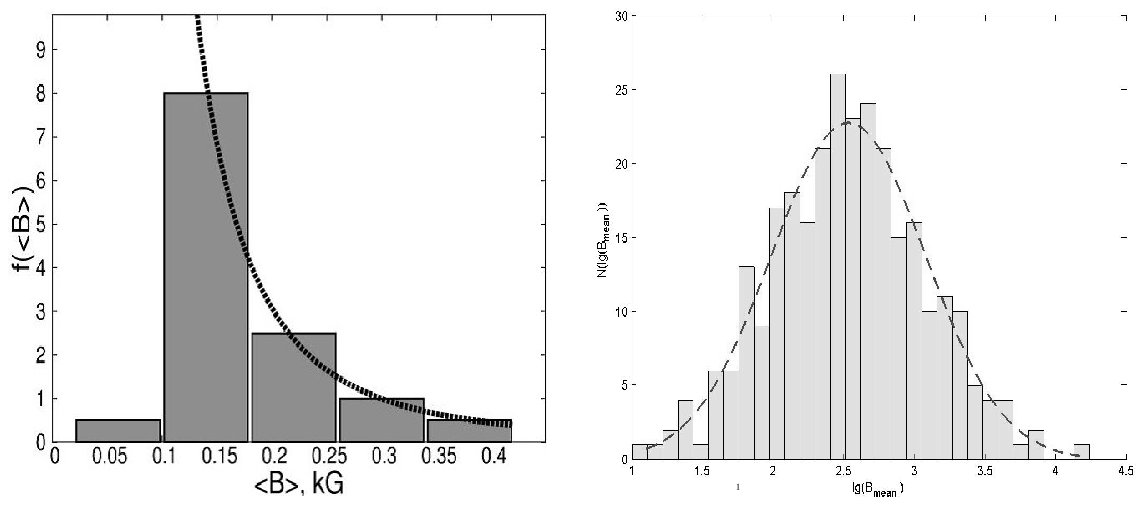}{Fig.MFdistrOBAstars} 
   {The distribution of the {\it rms} magnetic fields of known magnetic O stars (left panel) and
    the distribution of the {\it rms} magnetic field of BA stars in the catalogue by~\cite{Bychkov-2009} (right panel).
    }
%#########################################################################################################################==Figure1

The calculated by us function $f({\langle B\rangle})$ for all O stars with measured magnetic fields, 
including data by \citep{Hubrig-2011,Hubrig-2013} is given in Fig.~\ref{Fig.MFdistrOBAstars} (left panel) and can be 
fitted with a power law $f(\langle B\rangle) = A_0 \left(\frac{\langle B\rangle}{{\langle B\rangle}_0}\right)^\gamma \, ,$
where $A_0 =0.035$, $\gamma = -2.78$  for $\langle B\rangle > {\langle B\rangle}^{\mathrm{tr}} =100\,$G. 
%
% It means that the average magnetic fields of O-type stars are $8-9$ times weaker than those for BA stars.
%                                                      
The magnetic field function $f({\langle B\rangle})$ for BA stars was also fitted with a power law by \cite{Kholtygin-2015} for 
$\langle B\rangle > {\langle B\rangle}^{\mathrm{tr}} = 300\,G$. 
They obtained $A_0 =0.35 \pm  0.06$, and $\gamma = 2.09 \pm 0.13$. 
It means that the average magnetic fields of O-type stars are $8-9$ times weaker than those for BA stars.

At $\langle B\rangle <{\langle B\rangle}^{\mathrm{tr}}$ the magnetic field function greatly reduced relative 
to the power law. The relatively small number of stars with  $\langle B\rangle <{\langle B\rangle}^{\mathrm{tr}}$ was interpreted 
by \cite{Lignieres-2014} as an evidence of the {\it magnetic desert} in this field region as 
a result of the bifurcation between stable and unstable large scale magnetic field configurations. 

To clarify this issue we calculated the magnetic field distribution function using all suitable data from the 
catalogue by~\cite{Bychkov-2009}. We fit the magnetic field distribution by log--normal law instead of a power one. 
The result of our fit is presented in Fig.~\ref{Fig.MFdistrOBAstars} (right panel). 
We do not see the {\it magnetic desert} at least for BA stars.
The number of measured magnetic fields for O stars is to small to check if the log--normal law is better than 
the power one for these stars.

Recently \cite{Fossati-2015} created the histogram of the distribution of the dipolar magnetic field strength for the 
magnetic massive stars using their new measurements and did not detected the {\it magnetic desert} for intermediate-mass stars. 
They argued that the relatively weak fields might be more common than currently observed.

%============================================================================================================2.
\section{Magnetic field generation before the Main Sequence}
             
The population synthesis model of the magnetic field evolution for O and BA stars \citep{MedvedevKholtygin2015} 
showed that the present-day distribution of the magnetic fields  of OBA stars can be reproduced assuming 
that the initial magnetic field distribution at the ZAMS obeys the log--normal law.
For the sake of convenience the authors used the magnetic fluxes $F\approx 4\pi \left< B\right> R_*^2$ for stars with known 
values of $\left< B\right>$ instead of their magnetic fields. 
All obtained initial distributions of the magnetic fluxes at ZAMS were also fitted with log--normal law.
Recently  Medvedev \& Kholtygin (2016, this issue) show that parameters of the initial magnetic flux 
distribution can be chosen the same both for O and BA stars.  The mean logarithm of the initial magnetic fluxes at ZAMS for all 
OBA stars in the model without magnetic field dissipation (model I) is $\overline{F}=26.45$ with the standard deviation 
$\sigma = 0.50$, while for model II with a dissipation time $T_{\mathrm{d}} = 1/2 T_{\mathrm{MS}}$ 
$\overline{F}=26.87$ and $\sigma = 0.35$. Here $T_{\mathrm{MS}}$ is the star lifitime star on the~MS.

%###############################################################################################################################==Figure2
 \articlefigure[width=0.95\textwidth]{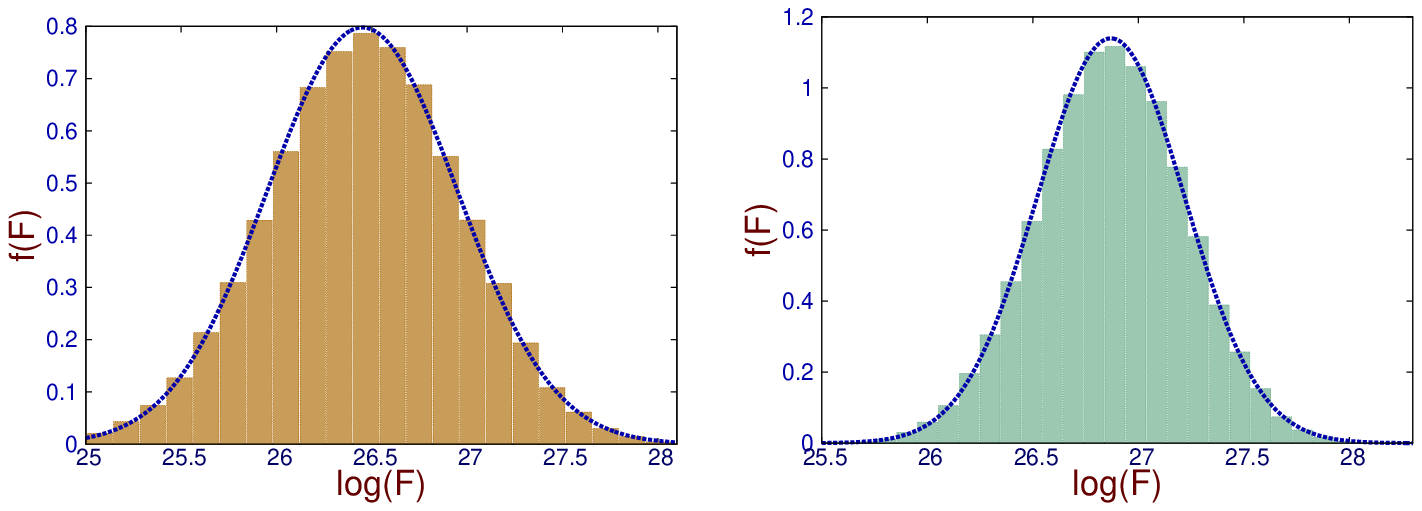}       {Fig.NFlux_ZAMS} 
        {Distribution of generated magnetic fluxes for model I (left panel) and model II (right panel) at the ZAMS. 
         Thick dashed lines show the initial magnetic flux distribution in the population synthhesis 
         model by~\cite{MedvedevKholtygin2015}.
         }
%###########################################################################################################################==Figure2_end

The nature of the log--normal distribution of magnetic fluxes for OB and BA stars is enigmatic.
 To explain this we can use the main idea by~\cite{Ferrario-2009} that mergers of protostars might play an important role in the 
formation of magnetic field of massive stars. 
Developing this idea, we assume that the magnetic field is not formed in a final merging of the protostars, but in multiple 
merging events of protostars and planetesimals. 

Suppose that there was $N$ cycles of merging before ZAMS and after act $i$ of merging the magnetic flux of star
$\left< F\right>_i = \alpha_i {\left< F\right>}_{i-1}$, where $\alpha_i$ is the coefficient of the field amplification 
and ${\left< F\right>}_{i-1}$ 
is the stellar magnetic flux at the cycle $i\!-\!1$. The amplification coefficients are supposed to be 
uniformly distributed in the interval $[a,b]$, where $a$ and $b$ are constant. 
The initial magnetic flux ${\left< F\right>}_{i=0} = F_0$ supposed to be identical for all stars. 
In Fig.~\ref{Fig.NFlux_ZAMS} we demonstrate the result of our modeling of the magnetic flux distribution at ZAMS. 
We fixed a value $N=20$ and fitted the parameters $a$, $b$ and $F_0$ only. The obtained optimal parameters are 
$a=1.0$, $b=2.49$, $F_0=7.9\cdot 10^{21}\,$G\,cm$^2$ for model I and $a=1.0$, $b=1.88$, $F_0=7.0\cdot 10^{23}\,$G\,cm$^2$ 
for model II. The total field amplification ratio $F_{N}/F_{0} \approx 3.5\cdot 10^4$ and $F_{N}/F_{0} \approx 10^3$ for models
I and II respectively.

%============================================================================================================2.
\section{Conclusions}

We show that the magnetic filed distribution for early BA-type stars can be described with the log--normal law. 
This means, in particular, the absence of a magnetic desert at least for early-type stars.
From our considerations we can conclude that multiple dynamo action during merging of protostars and planetesimals  
can be responsible for the generation of magnetic fields before the MS.

%_____________________________________________________________________________________________________________________

\acknowledgements AFK, SF, GAC, ASM, GGV and SVB thanks the RFBR grant 16-02-00604~A for a support. GGV also thanks for a 
                  support from the Russian Science Foundation (project 14-50-00043).

% \bibliography{refAFK}  % For BibTex

\begin{thebibliography}{}
 \expandafter\ifx\csname natexlab\endcsname\relax\def\natexlab#1{#1}\fi
 \expandafter\ifx\csname url\endcsname\relax
  \def\url#1{\texttt{#1}}\fi
 \expandafter\ifx\csname urlprefix\endcsname\relax\def\urlprefix{URL }\fi
 \providecommand{\eprint}[2][]{\url{#2}}

\bibitem[{{Bychkov} et~al.(2009){Bychkov}, {Bychkova}, \&
  {Madej}}]{Bychkov-2009}
{Bychkov}, V.~D., {Bychkova}, L.~V., \& {Madej}, J. 2009, \mnras, 394, 1338

\bibitem[{{Fabrika} et~al.(1997){Fabrika}, {Shtol'}, {Valyavin}, \&
  {Bychkov}}]{Fabrika-1997}
{Fabrika}, S.~N., {Shtol'}, V.~G., {Valyavin}, G.~G., \& {Bychkov}, V.~D. 1997,
  Astronomy Letters, 23, 43

\bibitem[{{Ferrario} et~al.(2009){Ferrario}, {Pringle}, {Tout}, \&
  {Wickramasinghe}}]{Ferrario-2009}
{Ferrario}, L., {Pringle}, J.~E., {Tout}, C.~A., \& {Wickramasinghe}, D.~T.
  2009, \mnras, 400, L71

\bibitem[{{Fossati} et~al.(2015){Fossati}, {Castro}, {Morel}, {Langer},
  {Briquet}, {Carroll}, {Hubrig}, {Nieva}, {Oskinova}, {Przybilla},
  {Schneider}, {Sch{\"o}ller}, {Sim{\'o}n-D{\'{\i}}az}, {Ilyin}, {de Koter},
  {Reisenegger}, \& {Sana}}]{Fossati-2015}
{Fossati}, L., {Castro}, N., {Morel}, T., {Langer}, N., {Briquet}, M.,
  {Carroll}, T.~A., {Hubrig}, S., {Nieva}, M.~F., {Oskinova}, L.~M.,
  {Przybilla}, N., {Schneider}, F.~R.~N., {Sch{\"o}ller}, M.,
  {Sim{\'o}n-D{\'{\i}}az}, S., {Ilyin}, I., {de Koter}, A., {Reisenegger}, A.,
  \& {Sana}, H. 2015, \aap, 574, A20. \eprint{1411.6490}

\bibitem[{{Hubrig} et~al.(2013){Hubrig}, {Sch{\"o}ller}, {Ilyin}, {Kharchenko},
  {Oskinova}, {Langer}, {Gonz{\'a}lez}, {Kholtygin}, {Briquet}, \& {Magori
  Collaboration}}]{Hubrig-2013}
{Hubrig}, S., {Sch{\"o}ller}, M., {Ilyin}, I., {Kharchenko}, N.~V., {Oskinova},
  L.~M., {Langer}, N., {Gonz{\'a}lez}, J.~F., {Kholtygin}, A.~F., {Briquet},
  M., \& {Magori Collaboration} 2013, \aap, 551, A33. \eprint{1301.4376}

\bibitem[{{Hubrig} et~al.(2011){Hubrig}, {Sch{\"o}ller}, {Kharchenko},
  {Langer}, {de Wit}, {Ilyin}, {Kholtygin}, {Piskunov}, {Przybilla}, \& {Magori
  Collaboration}}]{Hubrig-2011}
{Hubrig}, S., {Sch{\"o}ller}, M., {Kharchenko}, N.~V., {Langer}, N., {de Wit},
  W.~J., {Ilyin}, I., {Kholtygin}, A.~F., {Piskunov}, A.~E., {Przybilla}, N.,
  \& {Magori Collaboration} 2011, \aap, 528, A151. \eprint{1102.2503}

\bibitem[{{Kholtygin} et~al.(2010){Kholtygin}, {Fabrika}, {Drake}, {Bychkov},
  {Bychkova}, {Chountonov}, {Burlakova}, \& {Valyavin}}]{Kholtygin-2010a}
{Kholtygin}, A.~F., {Fabrika}, S.~N., {Drake}, N.~A., {Bychkov}, V.~D.,
  {Bychkova}, L.~V., {Chountonov}, G.~A., {Burlakova}, T.~E., \& {Valyavin},
  G.~G. 2010, Astronomy Letters, 36, 370. \eprint{1005.3705}

\bibitem[{{Kholtygin} et~al.(2015){Kholtygin}, {Hubrig}, {Valyavin}, {Fabrika},
  {Chuntonov}, {Dushin}, \& {Milanova}}]{Kholtygin-2015}
{Kholtygin}, A.~F., {Hubrig}, S., {Valyavin}, G.~G., {Fabrika}, S.~N.,
  {Chuntonov}, G.~A., {Dushin}, V.~V., \& {Milanova}, Y.~V. 2015, in Physics
  and Evolution of Magnetic and Related Stars, edited by Y.~Y. {Balega}, I.~I.
  {Romanyuk}, \& D.~O. {Kudryavtsev}, vol. 494 of Astronomical Society of the
  Pacific Conference Series, 221

\bibitem[{{Ligni{\`e}res} et~al.(2014){Ligni{\`e}res}, {Petit}, {Auri{\`e}re},
  {Wade}, \& {B{\"o}hm}}]{Lignieres-2014}
{Ligni{\`e}res}, F., {Petit}, P., {Auri{\`e}re}, M., {Wade}, G.~A., \&
  {B{\"o}hm}, T. 2014, in Magnetic Fields throughout Stellar Evolution, edited
  by P.~{Petit}, M.~{Jardine}, \& H.~C. {Spruit}, vol. 302 of IAU Symposium,
  338. \eprint{1402.5362}

\bibitem[{{Medvedev} \& {Kholtygin}(2015)}]{MedvedevKholtygin2015}
{Medvedev}, A., \& {Kholtygin}, A. 2015, in Physics and Evolution of Magnetic
  and Related Stars, edited by Y.~Y. {Balega}, I.~I. {Romanyuk}, \& D.~O.
  {Kudryavtsev}, vol. 494 of Astronomical Society of the Pacific Conference
  Series, 280

\end{thebibliography}
%
%
%
%
%

\end{document}